\documentstyle[prl,aps,floats,psfig]{revtex}
\draft
\newcommand{\nub}{{\overline{\nu}}}    
\newcommand{\ubar}{\overline{u}}    
\newcommand{\dbar}{\overline{d}}

\def\np#1#2#3   {{ Nucl. Phys.} {\bf#1}, #2 (#3). }
\def\pcps#1#2#3 {{ Proc. Cam. Phil. Soc.} {\bf#1}, #2 (#3). }
\def\pl#1#2#3   {{ Phys. Lett.} {\bf#1}, #2 (#3). }
\def\plc#1#2#3   {{ Phys. Lett.} {\bf#1}, #2 (#3); }
\def\prep#1#2#3 {{ Phys. Rep.} {\bf#1}, #2 (#3). }
\def\prev#1#2#3 {{ Phys. Rev.} {\bf#1}, #2 (#3). }
\def\prl#1#2#3  {{ Phys. Rev. Lett.} {\bf#1}, #2 (#3). }
\def\prs#1#2#3  {{ Proc. Roy. Soc.} {\bf#1}, #2 (#3). }
\def\ptp#1#2#3  {{ Prog. Th. Phys.} {\bf#1}, #2 (#3). }
\def\rmp#1#2#3  {{ Rev. Mod. Phys.} {\bf#1}, #2 (#3). }
\def\rpp#1#2#3  {{ Rep. Prog. Phys.} {\bf#1}, #2 (#3). }
\def\zp#1#2#3   {{ Zeit. Phys.} {\bf#1}, #2 (#3). }
\def\epj#1#2#3   {{ Eur. Phys. Jour.} {\bf#1}, #2 (#3). }

\begin{document}

\wideabs{
\title{Implication of $W$-boson
Charge Asymmetry Measurements in $p\overline{p}$ 
Collisions for Models of Charge Symmetry Violations
in Parton Distributions}

\author{ A.~Bodek$^{1}$, Q.~Fan$^{1}$, M.~Lancaster$^{2}$,
 K.S. McFarland$^{1}$, and U.~K.~Yang $^{1}$ }

\address{$^{1}$Department of Physics and Astronomy,
University of Rochester, Rochester, New York 14627 }
\address{$^{2}$Department of Physics and Astronomy,
 University College, London, UK}
%
%
%
\date{\today}
\maketitle
\begin{abstract}

A surprisingly
large charge symmetry violation of the sea quarks
in the nucleon has been proposed in a recent article
by Boros  {\em et al.} as an explanation of the discrepancy
between neutrino (CCFR) and muon (NMC) nucleon structure
function data at low $x$.  We show that these models are
ruled out by the published CDF $W$ charge asymmetry measurements,
which strongly constrain the ratio of
$d$ and $u$ quark momentum distributions 
in the proton over the $x$ range of $0.006$ to $0.34$. 
 This constraint also limits the systematic error from possible 
charge symmetry 
violation in the determinations of $\sin^2\theta_W$
from  $\nu N$ scattering experiments.
\end{abstract} 
\twocolumn
}

%
%
In a recent Physical Review Letter~\cite{BOROS},
Boros  {\em et al.}
proposed a model in which
a substantial charge symmetry violation (CSV) for parton
distributions in the nucleon accounts for the experimental
discrepancy between neutrino
(CCFR)~\cite{CCFR} and muon (NMC)~\cite{NMC}  nucleon structure
function data at low $x$. 
Charge symmetry (sometimes also referred to as isospin symmetry) 
is a symmetry which interchanges protons
and neutrons, thus simultaneously interchanging up and
down quarks, which implies the equivalence between the up (down) quark
distribution in the proton and the down (up) quarks in the neutron. 
Currently, all fits to Parton Distribution
Functions (PDFs) are preformed under the assumption of
charge symmetry between neutrons and protons.

Boros  {\em et al.} have proposed~\cite{BOROS} that charge symmetry
is broken such that  
the $d$ sea quark distribution
in the nucleon is larger than the $u$ sea quark distribution
for $x<0.1$, which also results in a violation of flavor symmetry. 
Their paper notes
that structure functions extracted in neutrino deep
inelastic scattering experiments are dominated by
the higher statistics data taken with neutrino (versus antineutrino)
beams. They note that neutrino-induced charged current interactions
couple to $d$ quarks and not to $u$ quarks, while the muon coupling to
the 2/3 charged $u$ quark is much larger than the coupling to
the 1/3 charge $d$ quark. Therefore,
if the $d$ sea quark distribution
is significantly larger than the $u$ sea quark distribution
in the nucleon, there would be a significant difference between
the nucleon structure functions as measured in neutrino and
muon scattering experiments. However, both neutrino and muon
scattering data have been taken on approximately isoscalar targets, 
such as iron or deuterium. 
Isoscalar targets have an equal number of neutrons and
protons. 
A larger number of $d$ sea quarks than $u$ sea quarks 
in an isoscalar target
implies a violation of charge symmetry. Therefore,
Boros {\em et al.} proposed  that a large charge symmetry 
violation  of the sea quarks in the nucleon might explain
the observed discrepancy $(10\sim15 \%)$ between neutrino and muon structure
function data.    

Boros  {\em et al.} define the following charge
symmetry violations in the nucleon sea.
\begin{eqnarray}
\delta \ubar (x) & = \ubar^p(x) - \dbar^n(x), \\
\delta \dbar (x) & = \dbar^p(x) - \ubar^n(x),
\label{eq:delta}
\end{eqnarray}
where $\ubar^p(x)$ and $\dbar^p(x)$
are the distribution of the 
 $u$ and $d$ sea anti-quarks in the proton, respectively.
Similarly $\ubar^n(x)$ and $\dbar^n(x)$
are the distribution of the
 $u$ and $d$ sea anti-quarks in the neutron, respectively.
The distributions for the quarks and antiquarks in
the sea is assumed to be the same. 
The relations for CSV in the sea quark distributions
are analogous to equations (1) and (2) for the sea anti-quarks.
Charge symmetry in the valence
quarks is assumed to be conserved, since there is good
agreement between the neutrino and muon scattering
data for $x>0.1$.

Within this model,
Boros {\em et al.} extract a large CSV
from the difference in structure functions
as measured in neutrino and muon scattering
experiments. 
 Theoretically, such a large charge
symmetry violation  (of order of 25\% to 50\%)
is very unexpected.
Therefore, the article has generated
a significant amount of interest
both within and outside 
the high energy physics community~\cite{science}.
 If the proposed model is valid,
all parametrizations of PDFs would have to be modified. In addition,
physics analyses which rely on the knowledge of PDFs (e.g. the
extraction of the electro-weak mixing angle from the ratio
of neutral current and charged current cross sections)  would be significantly
affected.

In this communication we show that the CSV models proposed
by Boros  {\em et al.}  are
ruled out by the $W$ charge asymmetry measurements
made by the CDF experiment at the Fermilab 
Tevatron collider~\cite{CDF}. These $W$ data provide a
very strong constraint on the ratio of
$d$ and $u$ quark momentum distributions 
in the proton over the $x$ range of $0.006$ to $0.34$.

\begin{figure}[t]
\centerline{\psfig{figure=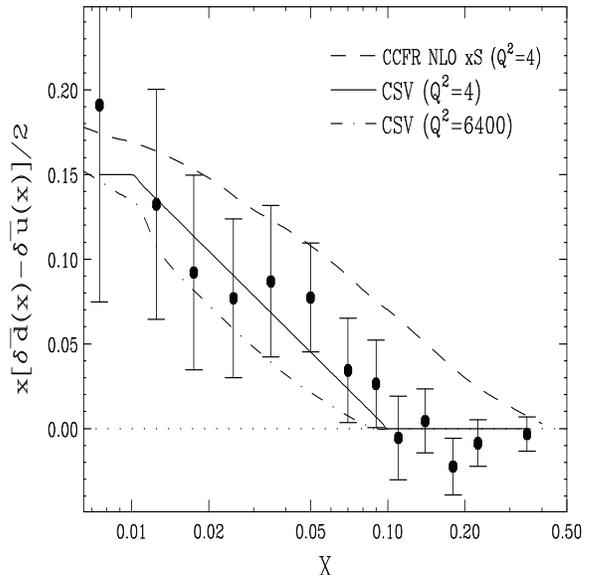,width=3.0in,height=3.0in}}
\caption{Charge symmetry violating distribution,
$x\Delta (x)$ = $x (\delta \dbar (x) - \delta \ubar (x))/2$  required
to explain the difference between neutrino and muon data,
as given in Fig. 3 of Boros {\it et al}.
The dashed line is the strange sea quark distribution
in the nucleon [$xs(x)$] as measured by the CCFR collaboration using
dimuon events produced in neutrino nucleon interactions.
 The solid line is our parametrization at $Q^2 =
4$ (GeV$/c$)$^2$ as described in the text. The dot-dashed line
is our parametrization when evolved to $Q^2 = M_W^2$.  }
\label{fig:DELTA} 
\end{figure}

Figure~\ref{fig:DELTA} shows the quantity 
$x\Delta (x) = x[\delta \dbar(x) - \delta \ubar(x)]/2$  required
to explain the difference between neutrino and muon data,
as given in Fig. 3 of Boros  {\em et al.}~\cite{BOROS}.
The average $Q^2$ of these data is about 4 (GeV$/c$)$^2$.
The dashed line is the strange sea quark distribution [$xs(x)$]
in the nucleon as measured by the CCFR collaboration using
dimuon events produced in neutrino nucleon interactions.
Boros  {\em et al.} state that 
the magnitude of implied charge symmetry violation is somewhere between
the full magnitude of the strange sea and half the magnitude
of the strange sea. Since the strange sea itself has been
measured to be about half of the average of the $d$ and $u$
sea, this implies a charge symmetry violation of
order 25\% (at $x=0.05$) and 50\% (at $x=0.01$).

However, as can be seen in Fig.~\ref{fig:DELTA}, the
shape of the strange sea does not provide a good
parametrization of the charge symmetry violation, therefore,
we have parametrized $x\Delta(x) $ (at $Q^2 =
4$ (GeV$/c$)$^2$ )   as follows. For $x > 0.1$,
$x\Delta(x) = 0.$  For $x < 0.01$,
 $x\Delta(x) =0.15$, and for
$0.01< x < 0.1$,
$x\Delta(x) = .15[log(x)-log(.1)]/[log(.01)-log(.1)]$.
This parametrization is shown as the solid line in
Fig.~\ref{fig:DELTA}.
The dot-dashed line shows the value of our parametrization
when evolved to  $Q^2 = M_W^2$. 
Boros  {\em et al.}
suggest that it is theoretically
expected that
$\Delta(x) = \delta \dbar (x) = - \delta \ubar (x)$,
which means that the sum of $u$ and $d$ sea distributions 
for protons and neutrons is the same. Within the assumption that
$\Delta(x) = \delta \dbar (x) = - \delta \ubar (x)$, we use
two models to parametrize the range of allowed changes in PDFs
to introduce the proposed charge symmetry violations.

In Model 1, it
is assumed that  the standard PDF parametrizations
are dominated by
neutrino data and therefore represent the average
of $d$ and $u$ sea quark distributions. Therefore,
half of the CSV 
is introduced into the $u$ sea quark distribution and
half of the effect is introduced into the $d$ sea
quark distribution such that the average of $d$ and $u$ sea
quark distributions is unchanged.
\begin{eqnarray}
\ubar^p(CSV) & = \ubar^p-\Delta(x)/2, \\
\dbar^p(CSV) & = \dbar^p+\Delta(x)/2, \\
\ubar^n(CSV) & = \ubar^n-\Delta(x)/2, \\
\dbar^n(CSV) & = \dbar^n+\Delta(x)/2. 
\label{eq:model1}
\end{eqnarray}
In Model 2, 
it is assumed that standard PDFs are dominated
by muon scattering data, and therefore are good
representation of the 2/3 charge $u$ quark distribution.
In this model,  the entire effect is introduced into the
$d$ sea quark distribution as follows; 
\begin{eqnarray}
\ubar^p(CSV) & = & \ubar^p, \\
\dbar^p(CSV) & = & \dbar^p+\Delta(x),\\
\ubar^n(CSV) & = & \ubar^n,\\
\dbar^n(CSV) & = & \dbar^n+\Delta(x).
\label{eq:model2}
\end{eqnarray}
Model 2 would change the total quark sea.


In order to have a precise test for the CSV effect, all PDFs 
have to be refitted based on the above two models. However, the ratio of $d$
and $u$ distribution will be almost
the same whether we refit the PDFs or not.
The $d/u$ ratio which has been extracted
from $F_2^n/F_2^p$ measurements (assuming charge symmetry) 
is in fact the quantity $u^n/u^p$  which does
not have any sensitivity to the proposed CSV effect.
In order to test for CSV effects, measurements of
$d^p/u^p$ or $d^n/u^n$  are required.
Therefore, the CDF measurements of the $W$ charge asymmetry
in $p\overline{p}$ collisions provide a unique test
of CSV effects, because of the direct sensitivity of these
data to the $d/u$ ratio in the proton (note that the $d$ and
$u$ quark distributions at small $x$ are dominated by
the quark-antiquark sea).
We now proceed to show that these implementations
of CSV in the nucleon sea are 
ruled out by the CDF $W$ charge asymmetry measurements
at the Tevatron.

At Tevatron energies, $W^+$ ($W^-$) bosons are produced in 
$p\overline{p}$ collisions primarily by the annihilation of 
$u$ ($d$) quarks in the proton  and $\overline{d}$ ($\overline{u}$) 
quarks from the antiproton. Because $u$ quarks carry on average 
more momentum than $d$ quarks~\cite{CTEQ4M},
the $W^+$ bosons tend to follow the direction
of the incoming proton and the $W^-$ bosons' that of the antiproton.
The charge asymmetry in the production of $W$ bosons as a function 
of rapidity ($y_W$) is therefore
related  to the difference in
the $u$ and $d$ quark distributions, and
is roughly proportional~\cite{ELB}~\cite{ADM}
 to the ratio of the difference and the sum of the
quantities $d(x_1)/u(x_1)$ and $d(x_1)/u(x_2)$, where
$x_1$ and $x_2$ are the fractions of the proton momentum carried
by the $u$ and $d$ quarks, respectively. (Note that
the quark distributions in the proton are equal to the
antiquark distributions in the antiproton). At large rapidity,
$x_1$ is larger than 0.1, which is a region
where CSV does not exist. On the other hand $x_2$
is in general less than 0.1, and a 25\% to 50\%
 CSV effect would imply
a very large effect on the $W$ asymmetry. Since the $W$ charge
asymmetry is sensitive to the $d/u$ ratio, it does not
matter  if the CSV effect at small $x$ is present in either $d$
or $u$ sea quark. All of these models would result in 
a similar change in the $W$ asymmetry.

\begin{figure}[t]
\centerline{\psfig{figure=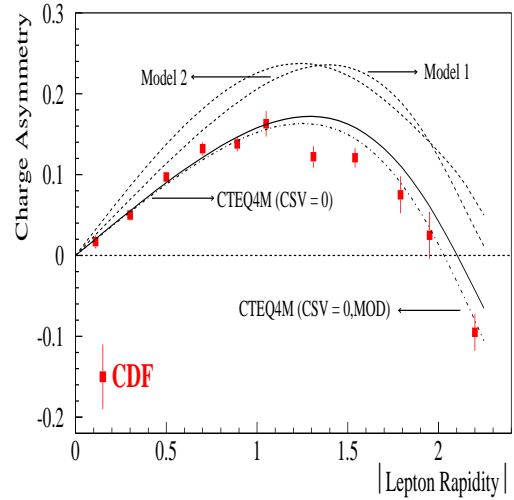,width=3.0in,height=3.0in}}
\caption{ The CDF $W$ Asymmetry data. The solid line is the
prediction from the standard CTEQ4M PDF(CSV=0).
The dashed-dotted line is the CTEQ4M PDF modified
for larger $d$ quark distribution at large $x$
as proposed by Yang and Bodek(CSV=0). The dashed and dotted
lines are predictions from the CTEQ4M PDF modified
to include the Boros  {\em et al.} charge symmetry violation
in the quark sea as described in the text. All theoretical predictions
are calculated in NLO QCD using the DYRAD program. }
\label{fig:WASYM}
\end{figure}

Experimentally, the $W$ rapidity is not determined
because of the unknown longitudinal momentum of the neutrino from the $W$ 
decay. What is actually measured
by the CDF collaboration
is the lepton charge asymmetry which is a convolution of 
the $W$ production charge asymmetry and the well known asymmetry from the
$V$-$A$ $W$ decay.
The two asymmetries are in opposite directions 
and tend to cancel at large values of rapidity.
 However, since the
$V$-$A$ asymmetry is well understood, the lepton asymmetry is still sensitive
to the parton distributions.
The lepton charge asymmetry is defined as:
\begin{equation}
A(y_l)=\frac{d\sigma^+/dy_l-d\sigma^-/dy_l}
            {d\sigma^+/dy_l+d\sigma^-/dy_l},
\end{equation}
where $d\sigma^+$ ($d\sigma^-$) is the cross section for
$W^+$~($W^-$) decay leptons as a function of lepton rapidity, with positive 
rapidity being defined in the proton beam direction.
The CDF data~\cite{CDF}  shown in
Fig.~\ref{fig:WASYM} span the broad range of lepton 
rapidity ($0.0<|y_l|<2.2$),
and provide  information about the $d/u$ ratio in the proton
over the wide $x$ range ($0.006<x<0.34$). 
Therefore, the CDF $W$ asymmetry data
would provide a strong tool to test the CSV models over a broad range
of $x$, and not just in part of the range proposed in the Boros {\em et al.}
model.
%

%
Also shown in Fig.~\ref{fig:WASYM} (solid line)
are the predictions for the $W$ asymmetry from QCD
calculated to Next-to-Leading-Order (NLO)
using the program DYRAD~\cite{Dyrad}, with 
the CTEQ4M PDF~\cite{CTEQ4M} parametrization
for the $d$ and $u$ quark distributions
in the proton ( we have used CTEQ4 
because it is the PDF set that has been used by Boros {\em et al.}
in their paper ).
As pointed out by Yang and Bodek~\cite{YANG},
the small difference between the data and the prediction of the
CTEQ4M PDF at high rapidity is because
the $d$ quark distribution is somewhat underestimated
at high $x$ in the standard PDF parametrizations.
The predictions of the CTEQ4M PDF with the proposed
modifications by Yang and Bodek are shown as the dashed-dotted
line in the figure.

The two dotted lines in 
Fig.~\ref{fig:WASYM} show
the predicted $W$ asymmetry for the
CTEQ4M PDF with the proposed Boros  {\em et al.} charge symmetry
violation in the sea for Model 1 and
Model 2, respectively. The
CDF $W$ data clearly rule out these models.
 
Most striking in this analysis is the broad range of lepton
rapidity over which this disagreement occurs with the
CSV models.  This is suggestive that models of this class
would be ruled out over a broad range of $x$, and not just
in part of the range proposed in the Boros  {\em et al.}
model.

In the direct measurement of the $W$ mass at the Tevatron,
the CDF $W$ asymmetry data have been used to limit the
error on $M_W$ from PDFs to about 15 MeV. This has
been done by calculating
 the  deviation between the error weighted
average measured asymmetry over the
rapidity range of the data, and the predictions from various
PDFs. This measured average asymmetry for
the data is  $0.087\pm0.003$. The predicted average asymmetries
(weighted by the same errors as the data) are 0.094, 0.125, and
0.141 for the CTEQ4M PDF, and for Model 1 and Model 2, respectively.
If we only accept PDFs which are within two standard deviations
of the CTEQ4M PDF, the $W$ asymmetry data rule out CSV effects
at the level of more than 10 standard deviations for the two models
with CSV effects.


Another precision measurement which is sensitive to CSV effects is the
measurement of neutral-current scattering in neutrino-nucleon collisions.
Just as the magnitude of the couplings to $u$ and $d$ quarks differ in
neutral-current $\mu$--$q$ scattering at NMC, the couplings
to $u$ and $d$ quarks also differ in
neutral-current $\nu$--$q$ scattering.  In this case, the left-handed and
right-handed couplings of the neutral current to quarks are given by
$g_L=I_3-Q\sin^2\theta_W$ and $g_R=-Q\sin^2\theta_W$, where $Q$ is the quark
charge and $I_3$ is the third component of the weak isospin in the quark
doublet, $+1/2$ for $u$-type quarks and $-1/2$ for $d$-type quarks.
Therefore the CSV-inspired enhancement in the
$d$ quark distributions will
change the the cross-section for neutral-current scattering, even
for an isoscalar target.  Because these cross-section
measurements are used to extract electroweak parameters, a CSV effect could
then affect the precision measurements of $\sin^2\theta_W$.

The most precise measurements of neutral-current neutrino-quark scattering
come from the CCFR~\cite{CCFR-NC} and NuTeV~\cite{NuTeV-NC} experiments.  As
noted above, CCFR had a beam of mixed neutrinos and anti-neutrinos, dominated
by neutrinos.  The NuTeV experiment
uses separate neutrino and anti-neutrino beams in its
measurements to allow separation of neutral-current neutrino and
anti-neutrino interactions.  The NuTeV and CCFR experiments 
measure combinations of the
ratios, $R^\nu$ and $R^\nub$, (above a fixed hadron energy
$\nu$
threshold of $20$~GeV
or $30$~GeV in NuTeV and CCFR, respectively), where
\begin{equation}
R^{\nu(\nub)}\equiv
\frac{\sigma_{\nu(\nub)}^{\rm NC}}{\sigma_{\nu(\nub)}^{\rm CC}}=
\frac{1}{2}-\sin^2\theta_W+\frac{5}{9}(1+r^{\pm 1} )\sin^4\theta_W,
\end{equation}
and $r\equiv\sigma_\nub^{\rm CC}/\sigma_\nu^{\rm CC}\sim0.4$.  The NuTeV 
experiment has
extracted $\sin^2\theta_W$ using the combination $R^\nu-rR^\nub$ which is
insensitive to the effects of sea quarks, and thus not changed by CSV effects
in sea, as in the Boros {\em et al.\,} model.  However, the CCFR measurement 
with a mixed beam 
is equivalent to $R^{\nu}+0.13R^{\nub}$ in which the sea quark
contributions do not cancel.

Within the framework of Model 1, the modified PDFs leave the charged
current neutrino data unchanged, but affect the level of
the neutral current cross section.
The effect of the Model 1 implementation of the Boros {\em et al.} 
model on the CCFR result has been calculated using the CTEQ4L
PDF~\cite{CTEQ4M} in the cross-section model.  The CCFR experiment 
extracts a
$\sin^2\theta_W$ which is equivalent to
$M_W=80.35\pm0.21$ GeV~\cite{CCFR-NC}, which can be compared to the current
average of all direct $M_W$ measurements, $80.39\pm0.06$ GeV.  Model 1 would
increase the CCFR measured $M_W$ by $0.26$~GeV.
Since the CDF $W$ asymmetry data rule out a CSV effect at the
the level of 1/5 of the magnitude of Model 1, the error from possible CSV
effects in PDFs is less than 50 MeV. This illustrates
the value of the CDF $W$ asymmetry
data in limiting the systematic error from PDF uncertainties
not only in the direct measurement of the $W$ mass in hadron colliders,
but also in the indirect measurement of the $W$ mass in neutrino
experiments.


In conclusion, the CDF $W$ asymmetry data rule out the Boros {\em et al.}
model for charge
symmetry violation in parton distributions~\cite{note} as the
source of the difference between neutrino (CCFR) and
muon (NMC) deep inelastic scattering data. Sources such as
a possible difference in nuclear effects between neutrino
and muon scattering, or a possible underestimate of the
strange quark sea in the nucleon have been ruled out~\cite{BOROS}. 
The experimental systematic errors between the two experiments,
and improved theoretical analyses of massive charm
production in both neutrino and muon scattering are both
presently being investigated~\cite{private}
as possible reasons for this discrepancy.
%


\begin{references}
\bibitem{note}
Boros  {\em et al.}
suggest that it is theoretically
expected that
$\Delta(x) = \delta \dbar (x) = - \delta \ubar (x)$,
which means that the {\it sum} of $u$ and $d$ sea distributions 
for protons and neutrons is the same. The CDF W Asymmetry data
indicate that the standard PDFs are correct for the proton
and cannot be modified. However, if this relationship is relaxed, 
all the charge and flavor symmetry violation can be put into the the PDFs
for the neutron only.  Preserving $F_2^n$  one can propose
the following relations:\\
$ \ubar^p(CSV)  =  \ubar^p$;
$ \ubar^n(CSV)  =  \ubar^n-2 \Delta(x)/5$;\\
$ \dbar^p(CSV)  =  \dbar^p$;
$ \dbar^n(CSV)  =  \dbar^n+8 \Delta(x)/5$.\\
In this model, the total neutron sea is much larger than
the proton sea, and there is a very large flavor symmetry
violation only in the neutron.
\end{references}
\end{document}